\documentclass[aps,pra,twocolumn,superscriptaddress,showpacs]{revtex4}

\usepackage{graphicx}

\begin{document}

\title{Optimal multicopy asymmetric Gaussian cloning of coherent states}

\author{Jarom\'{\i}r Fiur\'{a}\v{s}ek}
\affiliation{Department of Optics, Palack\'{y} University, 
17. listopadu 50, 77200 Olomouc, Czech Republic}

\author{Nicolas J. Cerf\,}
\affiliation{QuIC, Ecole Polytechnique, CP 165, 
Universit\'{e} Libre de Bruxelles, 1050 Brussels, Belgium }

\begin{abstract}
We investigate the asymmetric Gaussian cloning of coherent states
which produces $M$ copies from $N$ input replicas, such that the fidelity
of all copies may be different. We show that the optimal asymmetric 
Gaussian cloning can be performed with a single phase-insensitive amplifier
and an array of beam splitters. We obtain a simple analytical expression 
characterizing the set of optimal asymmetric Gaussian cloning machines. 
\end{abstract}


\maketitle

\section{Introduction}

The perfect copying of unknown quantum states is forbidden by the linearity of quantum
mechanics \cite{Wootters82}. This observation lies at the heart of novel quantum
communication protocols, such as quantum key distribution (QKD) which allows the provably
secure sharing of a secret key between two distant partners (see, e.g., \cite{Gisin02}). 
Any eavesdropping on a QKD system introduces noise into the transmission, 
which can be detected by the legitimate users. The optimal individual eavesdropping 
attacks on many QKD protocols consist of the optimal (approximate) copying of the quantum 
states transmitted in the channel, where one copy is sent to the legitimate receiver 
while the other copy is kept by the eavesdropper and measured upon at a later stage. 
After the seminal paper by Bu\v{z}ek and Hillery  \cite{Buzek96}, where 
the concept of universal quantum cloning machine was introduced, the issue of
quantum copying has attracted considerable attention (see, e.g., \cite{rev1,rev2}).
This effort culminated in the recent years with the experimental demonstration of optimal 
$1 \rightarrow 2$ cloning machines for polarization states of photons  
based either on parametric amplification \cite{Lamas-Linares02,Fasel02,deMartini04} 
or on the symmetrization of the multiphoton state 
on an array of beam splitters \cite{Ricci04,Irvine04}. The latter technique 
was also exploited lately to realize the universal symmetric cloning machine 
for qubits that produces three clones \cite{Masullo04}.

In applications such as QKD, one is often interested in asymmetric cloning where
the fidelities of the clones are different. This is indeed necessary to study 
the trade-off between the information gained by the eavesdropper 
and the noise detected by the legitimate users.
The optimal $1\rightarrow 2$ asymmetric cloning of qubits and qudits has been studied in
detail \cite{Cerf98,Niu98,Cerf00} and very recently an experimental demonstration 
of $1\rightarrow 2$ asymmetric cloning of polarization states of photons \cite{Zhao04} 
based on partial teleportation \cite{Filip04} was reported. Going beyond two copies,
multipartite asymmetric cloning machines have been  introduced in \cite{Iblisdir04},
which produce $M$ copies with different fidelities $F_j$ ($j=1,\cdots M$).
Several examples of such multipartite asymmetric cloners for qubits 
and qudits were presented in \cite{multiasymm1,multiasymm2}.

In the context of the rapid development of quantum information processing with continuous
variables \cite{Braunstein04}, the cloning of coherent states has been extensively studied 
over the last years \cite{Lindblad00,Cerf00CV}. 
It was shown that the optimal $N\rightarrow M$ symmetric 
cloning of optical coherent states that preserves the Gaussian shape of the Wigner function 
can be accomplished with the help of a phase-insensitive amplifier followed by an
array of beam splitters that distributes the amplified signal 
into $M$ modes \cite{Braunstein01,Fiurasek01}.
It is also possible to exploit the off-resonant interaction of light beams with atomic
ensembles and perform the cloning of coherent states into an atomic memory \cite{Fiurasek04}. 
The cloning of a finite distribution of coherent states was studied \cite{Cochrane04}, 
and a reversal of cloning by means of local operations and classical 
communication was suggested in \cite{Filip04CV}.

On the experimental side, the $1\rightarrow 2$ optimal Gaussian cloning 
of coherent states of light was recently demonstrated in \cite{Andersen05}. 
There, the phase-insensitive optical amplifier was replaced with a clever combination of
beam splitters, homodyne detection, and feedforward, which effectively simulated the
amplification process. Using homodyne detectors with very low electronic noise, it was
possible to achieve a cloning fidelity of about $65$\%, very close to the theoretical
maximum of $2/3$. In another experiment, the $1\rightarrow 2$ telecloning of 
coherent states of light was also realized \cite{Koike06}.

In this paper, we extend the concept of multipartite asymmetric cloning to continuous
variables and present the optimal multipartite asymmetric Gaussian cloning machines 
for coherent states.  These devices produce $M$ approximate replicas 
of the coherent state $|\alpha\rangle$ from $N$ input replicas, such that 
the fidelity $F_j$ of each clone is generally different, and, for a given set
of $F_1, \ldots,F_{M-1}$, the fidelity of the $M$-th clone $F_M$ is the maximum possible.
The multicopy asymmetric $1 \rightarrow M$ cloning of coherent state was previously studied 
by Ferraro and Paris in the context of telecloning \cite{Ferraro05}. Here, 
we rigorously prove that their scheme is optimal, and present a generic optimal asymmetric  cloning machine for any number $N$ of input replicas, as well as its optical implementation. 
We also consider the related problem of the optimal partial state estimation 
of coherent states.

The rest of the paper is structured as follows. In Section~II, we present an optical
cloning scheme based on phase-insensitive amplification and passive linear optics. We also
derive the trade-off between the fidelities (or, equivalently, added thermal noises), which
fully characterizes the class of the optimal multipartite asymmetric Gaussian cloners. 
In Section~III, we describe an alternative cloning scheme where the  amplification 
is replaced by measurement and feedforward. In Section~IV, we shortly discuss 
the relationship between the optimal asymmetric cloning and optimal partial measurement 
of coherent states. Then, the proof of the optimality of the asymmetric cloning machine
is given in Section~V. Finally, Section~VI contains a brief summary and conclusions.

\section{Asymmetric Gaussian cloning of coherent states}

In what follows, we restrict ourselves to Gaussian cloning transformations. Note that
it has been found in \cite{Cerf04} that the optimal $1 \to 2$ cloning transformation 
for coherent states (i.e. the transformation that maximizes the single-clone fidelity) 
is in fact non-Gaussian, though the gain in fidelity is tiny. 
Nevertheless, it should be stressed that, while this non-Gaussian cloner 
is optimal in terms of fidelity, it adds more noise to the clones (as measured by 
the quadrature variances) than the optimal Gaussian cloner. 
In potential applications of cloning such as eavesdropping 
on QKD protocols with coherent states and homodyne detection \cite{Grosshans02,Grosshans03}, 
one is often interested in minimizing the quadrature variance. In such a case,
the Gaussian cloning turns out to be the most dangerous attack \cite{Grosshans04}. 
Finding the multipartite generalization of asymmetric Gaussian cloning is 
therefore a very interesting question.

As we will prove in Section~V, the optimal Gaussian cloning machine 
has the simple structure depicted in Fig.~1, which is a direct generalization 
of the $1 \rightarrow 2$ asymmetric cloner \cite{Fiurasek01}. 
The signal contained in $N$ input replicas of the coherent state $|\alpha\rangle$
is first collected into a single mode by an array of $N-1$ unbalanced beam 
splitters \cite{Braunstein01,Fiurasek01}.
After this, a single mode $a$ carries all the signal, and is in a coherent state
$|\sqrt{N}\alpha\rangle$. This mode is sent on an unbalanced beam splitter BS with 
amplitude transmittance $t$ and reflectance $r$, 
which divides the signal into two modes $a$ and $b_1$. 
Mode $a$ is then amplified in a phase-insensitive amplifier (NOPA) with amplitude gain $g$.
The modes $a$ and $b_1$ together with $M-2$ auxiliary modes $b_j$, 
with $j=2,\ldots, M-1$,
are combined in a passive linear $M$-port interferometer IF whose output modes contain the
$M$ clones. The interferometer is designed in such a way that that the coherent component 
in each output mode is equal to $\alpha$. In the Heisenberg picture, 
the overall input-output transformation describing the cloner depicted in Fig.~1 reads
\begin{equation}
a_j=\frac{1}{\sqrt{N}}\, a+\sum_{k=1}^{M-1} \kappa_{jk} \, b_k +\sqrt{n_j} \, c^\dagger,
\label{cloningtransformation}
\end{equation}
where $c^\dagger$ is the creation operator of the idler port of the amplifier
and $b_k$ are the annihilation operators of the $M-1$ auxiliary modes, initially 
in the vacuum state. Here, $n_j$ represents the amount of noise added 
to the $j$-th clone, 
and the $\kappa_{jk}$ coefficients are chosen in such a way that the canonical commutation relations are conserved.

\begin{figure}[!t!]
\begin{center}
\includegraphics[width=\linewidth]{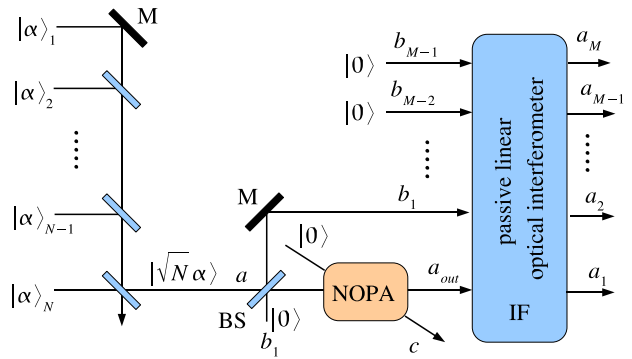}
\end{center}
\caption{Optimal Gaussian $N\rightarrow M$ fully asymmetric cloning of coherent states.
See text for details. } 
\end{figure}

It follows from the canonical transformations (\ref{cloningtransformation}) that each clone 
is in a mixed Gaussian state with coherent amplitude $\alpha$ and added thermal noise characterized by a mean number of thermal photons $n_j$. 
The Husimi $Q$-function of the $j$-th clone reads,
\begin{equation}
Q_j(\beta)=\frac{1}{\pi(n_j+1)} \exp\left(-\frac{|\alpha-\beta|^2}{n_j+1}\right).
\label{Qfunction}
\end{equation}
The fidelity of the $j$-th clone is proportional to the value of the Husimi 
Q-function at $\beta=\alpha$, and is therefore a monotonic function of the added
thermal noise, 
\begin{equation}
F_j=\frac{1}{1+n_j}.
\label{fidelity}
\end{equation}
The cloning is covariant and isotropic, i.e., the fidelity does not depend on
the input state $|\alpha\rangle$ and the added noise is the same for each quadrature.
These are natural conditions that the optimal cloning machine should satisfy.

The shot-noise limited amplification is governed by the transformation
\begin{equation}
a_{\mathrm{out}}=g\, (t\, a-r\, b_1)+\sqrt{g^2-1}\, c^{\dagger}
\label{amplification}
\end{equation}
and the total mean number of  thermal photons produced during the amplification is
$n_{\mathrm{tot}}=g^2-1$. Since the linear interferometer does not add any noise, we have
\begin{equation}
g=\sqrt{1+n_{\mathrm{tot}}},
\label{gain}
\end{equation}
where $n_{\mathrm{tot}}=\sum_{j=1}^M n_j$.
The total intensity of the coherent signal after amplification is 
$N(r^2+g^2t^2)|\alpha|^2 $, which should be equal to $M|\alpha|^2$
if we require the coherent component of each clone to be equal to $\alpha$. 
From this, we can determine the transmittance of BS, namely
\begin{equation}
t=\sqrt{\frac{M-N}{n_{\mathrm{tot}}\, N}}.
\label{transmittance}
\end{equation}
The multiplets of $n_j$ cannot be arbitrary. Indeed, the $M$-port interferometer IF 
in Fig.~1  is described by  a unitary matrix $V$, such that 
$a_j=\sum_{j=1}^{M-1} v_{jk} b_k+v_{jM} a_{\mathrm{out}}$. 
The unitarity of $V$ imposes a constraint on $n_j$ which can be expressed as 
\begin{equation}
\left(\sum_{k=1}^M \sqrt{n_k}\right)^2 = (M-N)\left(\sum_{j=1}^M n_j +1\right).
\label{noisecondition}
\end{equation}
This formula provides a simple analytical parametrization of the set of optimal
$N\rightarrow M$ multipartite asymmetric Gaussian cloning machines for coherent states.

In the special case of a $1\to 2$ asymmetric Gaussian cloner, Eq.~(\ref{noisecondition})
reduces to
\begin{equation}
n_1 \, n_2 = (1/2)^2  \; ,
\end{equation}
which coincides with the no-cloning uncertainty relation 
that was displayed in \cite{Cerf00CV,Fiurasek01}. 
(Note that 1/2 corresponds here to one shot-noise unit.)
Interestingly, if we consider a $1\to 3$ cloner and assign to the first two clones
the fidelity of the optimal $1\to 2$ symmetric cloner, that is, $n_1=n_2=1/2$, 
we obtain by solving Eq.~(\ref{noisecondition}) that the noise of the third clone
is not infinite, $n_3=2$. 
As noticed in \cite{multiasymm1}, this means that some quantum information remains available
beyond the one contained in the two clones (it actually corresponds to the information
hidden in the anticlone).
In the case where $N=1$ but $M$ is arbitrary, we recover the expression 
that was derived by Ferraro and Paris \cite{Ferraro05}.
Finally, note that if one clone is perfect, e.g., $n_M=0$, then Eq.~(\ref{noisecondition})
is transformed into the same equation but for a $(N-1) \to (M-1)$ cloner, which means
that one of the input replicas is simply redirected to the perfect clone while the
cloning of the $N-1$ remaining input replicas into the $M-1$ other clones is simply
governed by the same relation.

\section{Optimal asymmetric cloning via measurement and feedforward} 

In the experimental demonstration of the optimal cloning
of coherent states of light carried out in \cite{Andersen05},
the amplification in a phase-insensitive amplifier was replaced
by a clever combination of a partial measurement and feedforward 
that effectively simulates the amplification process. 
Since an amplifier of arbitrary gain $g$ can be implemented in this
way \cite{Josse06}, the scheme shown in Fig.~1 can be straightforwardly 
transformed into a setup which involves only passive linear optics, 
balanced homodyne detection and coherent displacement
of the beams proportional to the measurement outcomes. The resulting configuration is
illustrated in Fig.~2.

\begin{figure}[!t!]
\begin{center}
\includegraphics[width=\linewidth]{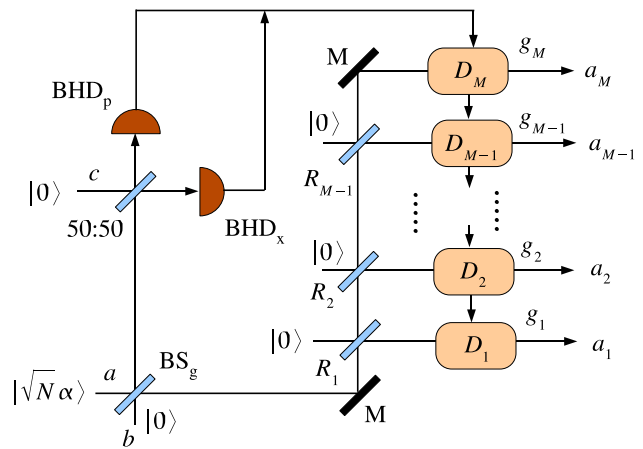}
\end{center}
\caption{Setup for the multipartite asymmetric Gaussian cloning of coherent states 
using homodyne detection and feedforward.}
\end{figure}

We assume that all available signal has been collected into a single
mode $a$ which is thus in the coherent state $|\sqrt{N}\alpha\rangle$. 
The beam is divided into
two parts on a beam splitter BS$_g$ with amplitude transmittance $\tilde{t}$ 
and reflectance $\tilde{r}$. The reflected part 
is fed into a heterodyne detector consisting of a balanced beam splitter 
whose auxiliary input port $c$ is in the vacuum state and two balanced 
homodyne detectors BHD measuring the $x$ and $p$ quadratures, respectively. 
This detector effectively measures the operator 
$o=\tilde{r}\, a+\tilde{t}\, b+c^\dagger$. The portion of the beam transmitted through BS$_g$,
characterized by $\tilde{t}\, a-\tilde{r}\, b$, is divided into $M$
modes $a_j$ by an array of $M-1$ beam splitters with reflectances $r_k$. Each 
mode $a_k$ is then coherently displaced by amount $g_j \, o$, where $g_j$ is
the electronic gain of the corresponding feed forward. The added thermal noise in the
output mode $a_j$ reads $n_j=g_j^2$, which immediately fixes all electronic gains,
\begin{equation}
g_j=\sqrt{n_j}.
\end{equation}
As shown in \cite{Josse06}, the optical amplification gain $g$ is obtained 
in the feed-forward equivalent scheme if the beam is split on a beam splitter 
with reflectance $\sqrt{1-1/g^2}$, and the reflected part is heterodyne measured.
Since in the scheme of Fig.~1 only an (amplitude) fraction $t$ of 
the input beam $a$ is actually amplified, we see that by replacing the amplifier 
with a beam splitter of reflectance $\sqrt{1-1/g^2}$, the fraction of the beam
that is sent to the heterodyne detector is $t\,\sqrt{1-1/g^2}$.
Thus, in Fig.~2, the reflectance  of the beam splitter 
BS$_g$ must be $\tilde{r}=t\, \sqrt{1-1/g^2}$, which yields
\begin{equation}
\tilde{r}=\sqrt{\frac{M-N}{(1+n_{\mathrm{tot}})\,N}}.
\end{equation}
It remains to determine the reflectances $r_k$ of the final array of beam splitters.
They are fixed by the condition that the coherent amplitude of each clone is $\alpha$.
After some algebra we find that 
\begin{equation}
r_j = 
\frac{\sqrt{1+n_{\mathrm{tot}}}-\sqrt{(M-N)\,n_j}}{\sqrt{(2+n_{\mathrm{tot}})N-M}} \, 
\prod_{k=1}^{j-1}(1-r_k^2)^{-1/2}.
\label{recurs}
\end{equation}
From this formula, all $r_k$'s can be calculated in an iterative way, 
starting from $r_1$ [which is given by Eq.~(\ref{recurs}) for $j=1$
with the product over $k$ replaced by 1].

\section{Optimal cloning and optimal partial estimation of coherent states}

There is a close relationship between optimal cloning and optimal state estimation. 
An interesting scenario that recently attracted a lot of attention consists in
the partial estimation of a state, which yields the classical estimate of the
state as well as the perturbed quantum state \cite{Banaszek01,Mista05,Mista06b,Sacchi06,Sciarrino06,Andersen06}. 
According to the fact that in the limit of an infinite number of copies, the optimal cloning
becomes equivalent to optimal state estimation \cite{Bae06}, this optimal
partial estimation can be viewed as a limiting case of an asymmetric cloning producing one 
(quantum) copy with fidelity $F$ and infinitely many (classical) copies with 
fidelity $G$ \cite{Iblisdir04}. From the analytical formula (\ref{noisecondition}),
we can thus rigorously derive the optimal trade-off between the fidelities $F$ and $G$ 
in the partial Gaussian estimation of coherent states. 
We set $n_1=n_F$, $n_2=n_3=\ldots =n_M=n_G$ and take the limit $M \rightarrow \infty$, 
which results in
\begin{equation}
n_G=\frac{(n_F+1)^2}{4n_F}.
\label{eq-trade-off}
\end{equation}
In the limit of an undisturbed quantum copy ($n_F=0$), we have an infinitely
noisy state estimation, as expected. We also note that $n_F=n_G=1$ is a solution
of Eq.~(\ref{eq-trade-off}), which corresponds to the optimal (full) estimation 
of coherent states. Equation~(\ref{eq-trade-off}) also translates in the following
relation between the fidelities,
\begin{equation}
G=\frac{4F(1-F)}{4F(1-F)+1}.
\end{equation}
This agrees with the trade-off derived in \cite{Andersen06} which confirms that the experimentally demonstrated partial measurement of coherent states 
in that work was indeed optimal among all Gaussian
strategies. Note that using a non-Gaussian protocol a slightly better trade-off between
$F$ and $G$ could be achieved \cite{Mista06}.

\section{Proof of optimality}

In what follows, we will prove the optimality of the asymmetric cloner defined
in Secs.~II and III. Let us first note that the fidelities 
are monotonic functions of $n_j$,
so that instead of maximizing the fidelities $F_j$ we can equivalently minimize the
added thermal noise $n_j$. 
The design of the optimal cloner can be thus rephrased as the minimization of a
cost function \cite{Iblisdir04} 
\begin{equation}
C(n_j)=\sum_{j=1}^M x_j n_j,
\label{costfunction}
\end{equation}
which is a linear convex mixture of $n_j$, $x_j > 0$.  The ratios of the
coefficients $x_j$ control the asymmetry of the cloning machine. 

The most general Gaussian operation is a trace-preserving 
Gaussian completely positive (CP) map \cite{Lindblad00}, and we have 
to minimize (\ref{costfunction}) over all such maps. 
At the level of covariance matrices  $\gamma$, the Gaussian CP map acts as
\begin{equation}
\gamma_{\mathrm{out}}=S \gamma_{\mathrm{in}} S^T +G.
\label{GaussianCPmap}
\end{equation}
The covariance matrix of $N$ modes is defined as $\gamma_{jk}=\langle \Delta r_j \Delta
r_k+\Delta r_k\Delta r_j\rangle$, where $r=(x_1,\ldots,x_N,p_1\ldots,p_N)$ is
the vector of quadrature operators, $[x_j,p_k]=i\delta_{jk}$. The first moments
transform under the Gaussian CP map 
according to $\langle r_{\mathrm{out}}\rangle=S \langle r_{\mathrm{in}}\rangle$.

The matrices $S$ and $G$ must satisfy the complete positivity constraint
\begin{equation}
A \equiv G + i K  \geq 0 \, , \qquad K = J_{M_{\mathrm{out}}} - S J_{M_{\mathrm{in}}} S^T \, ,
\label{completepositivity}
\end{equation}
where the matrix
\begin{equation}
i J_M= i \left(
\begin{array}{cc}
0 & I \\
-I & 0  \\
\end{array}
\right),
\label{Jmatrix}
\end{equation}
comprises the commutators of the quadrature operators, while
$I$ denotes the identity matrix of dimension $M$, and $M_{\mathrm{in}}$ and $M_{\mathrm{out}}$ are
the number of input and output modes, respectively.
The cloning machine of interest has effectively only a single input mode $a$ (as
we collect the $N$ input signals into a single mode) and $M$ output modes, so that
$M_{\mathrm{in}}=1$ and $M_{\mathrm{out}}=M$. The $M \times 2$ matrix $S$ is fixed by the condition 
that the first moments should be preserved by cloning. We get
\begin{equation}
S^T=\frac{1}{\sqrt{N}}\left(
\begin{array}{cccccccc}
1 & 1 & \ldots & 1 & 0 & 0 & \ldots & 0 \\
0 & 0 & \ldots & 0 & 1 & 1 & \ldots & 1 \\
\end{array}
\right).
\label{Smatrix}
\end{equation}

We have to minimize $C(n_j)$ over the set of all Gaussian completely positive 
maps (\ref{GaussianCPmap}) with the matrix $S$ given by (\ref{Smatrix}), that is, 
we have to optimize over all $G$'s satisfying (\ref{completepositivity}). 
Clearly, if $G_1$ and $G_2$ satisfy (\ref{completepositivity}), then any convex
combination $pG_1+(1-p)G_2$ with $p\in [0,1]$ also does. 
Thus, we have a convex optimization problem.
Moreover, since $n_j=(G_{jj}+G_{M+j,M+j}+2/N-2)/4$,
the cost function (\ref{costfunction}) is linear in the matrix elements of $G$.
Hence, the problem amounts to minimizing 
\begin{equation}
\tilde{C}(G)=\sum_{j=1}^M x_j (G_{jj}+G_{M+j,M+j})
\label{noisecostfunction}
\end{equation}
under the constraints (\ref{completepositivity}), which is an instance of a linear
semidefinite program \cite{Boyd96}. We shall now prove the optimality of
(\ref{cloningtransformation}) by deriving a lower bound on $\tilde{C}(G)$ which 
is saturated by (\ref{cloningtransformation}).

The specific feature of the transformation (\ref{cloningtransformation}) 
is that only a single creation operator $c^\dagger$ is admixed to the annihilation 
operators. This operator is responsible for the added noise in cloning. 
Since all modes are initially in coherent states, the normally ordered moments 
of the operators $a_j$ for the clones can be easily calculated,
\begin{eqnarray*}
&\langle \Delta a_j \Delta a_k \rangle= \langle \Delta a_j^\dagger \Delta
a_k^\dagger\rangle =0, & \\
&\langle \Delta a_j^\dagger \Delta a_k \rangle=\sqrt{n_j n_k}.&
\end{eqnarray*}
The covariance matrix of the $M$ clones is then fully determined by the added noises $n_j$,
\begin{equation}
\gamma_{\mathrm{out}}=\left(
\begin{array}{cc}
I+2F & 0 \\
0 & I+2F
\end{array}
\right),
\label{gammaout}
\end{equation}
where $F$ is a symmetric $M \times M$ matrix with elements $F_{jk}=\sqrt{n_j n_k}$.
Since the matrix $S$ is fixed and the input state of the cloner is  a coherent state with
covariance matrix $\gamma_{\mathrm{in}}=I$, the matrix $G_{\mathrm{opt}}$ corresponding to transformation
(\ref{cloningtransformation}) can be determined from Eq. (\ref{GaussianCPmap}). This yields
\begin{equation}
G_{\mathrm{opt}}=\left(
\begin{array}{cc}
I+2F-\frac{1}{N}H & 0 \\
0 & I+2F-\frac{1}{N} H
\end{array}
\right),
\label{Gmatrix}
\end{equation}
where $H$ is a matrix whose elements are all equal to one, $H_{jk}=1$.

Since the transformation (\ref{cloningtransformation}) can be associated
with a CP map, the matrix $G_{\mathrm{opt}}$ must satisfy the inequality (\ref{completepositivity}).
For the particular $S$ matrix (\ref{Smatrix}), this gives
\begin{equation}
A_{\mathrm{opt}} \equiv \left(
\begin{array}{cc}
I+2F-\frac{1}{N}H & i(I-\frac{1}{N}H) \\[1mm]
-i(I-\frac{1}{N}H) & I+2F-\frac{1}{N} H \\
\end{array}
\right) \geq 0
\label{Amatrix}
\end{equation}
We can transform the matrix $A_{\mathrm{opt}}$ to a block-diagonal form with the help of the unitary matrix
\begin{equation}
U=\frac{1}{\sqrt{2}}\left(\begin{array}{cc}
I & iI \\[1mm]
iI & I
\end{array}
\right)  \, ,
\label{Umatrix}
\end{equation}
which gives $U A_{\mathrm{opt}} U^\dagger =\mathrm{diag}(2F+2I-2H/N,2F)$, so that there remains
to prove that $F+I-H/N \geq 0$ and $F\geq 0$.
The matrices $F$ and $H$ both have rank one, and can be written in Dirac notation as 
$F= |f\rangle \langle f|$, where $|f\rangle=\sum_{j=1}^M \sqrt{n_j}|j\rangle$, and
$H=|h\rangle\langle h|$, where $|h\rangle=\sum_{j=1}^M |j\rangle$. The condition $F \geq
0$ is satisfied by definition, while solving $F+I-H/N \geq 0$ yields the nontrivial 
constraint on $n_j$'s. Actually, it is sufficient to check the latter positivity condition 
in the two-dimensional subspace spanned by the (un-normalized) vectors $|f\rangle$ 
and $|h\rangle$, and one can prove that it is indeed satisfied 
provided that Eq.~(\ref{noisecondition}) holds. 

We now derive a tight lower bound on $\tilde{C}(G)$, which is saturated by
(\ref{cloningtransformation}).  Suppose that we find a positive
semidefinite matrix $Z \geq 0$ such that it satisfies the conditions
\begin{equation}
\mathrm{Tr}[ZG]=\tilde{C}(G)
\label{Zcost} 
\end{equation}
and 
\begin{equation}
Z\, A_{\mathrm{opt}} \equiv Z(G_{\mathrm{opt}}+iK)=0,
\label{Ztrace}
\end{equation}
where $iK=iJ_{M} -i S J_{1} S^T$. Then, the Gaussian CP map with matrix $G_{\mathrm{opt}}$ 
is the optimal one that minimizes $\tilde{C}(G)$. 
Since for every admissible  $G$ we have $G+iK \geq 0$, it follows from $Z\geq 0$ that 
$\mathrm{Tr} [Z(G+iK)] \geq 0$, which implies that 
$\tilde{C}(G)\geq -i \mathrm{Tr}[ZK]$, $\forall G$.
Equation (\ref{Ztrace}) implies that this lower bound is saturated by $G_{\mathrm{opt}}$, which 
is therefore optimal.

The matrix $Z$ can be determined from Eqs. (\ref{Zcost}) and (\ref{Ztrace}). We find that
\begin{equation}
Z=\left(\begin{array}{cc}
X & iY \\[1mm]
-iY & X
\end{array}
\right) ,
\label{Zmatrix}
\end{equation}
where $X=\mathrm{diag}(x_1,\ldots,x_M)$ is fixed by Eq.~(\ref{Zcost}),
while $Y$ is a real symmetric matrix that satisfies
\begin{eqnarray}
&&Y(I-N^{-1}H)+X(I+2F-N^{-1} H)=0 \, , \nonumber \\
&&XF=YF \, ,
\label{Yconstraints}
\end{eqnarray}
as a consequence of Eq.~(\ref{Ztrace}). Note that $X>0$ by definition because $x_j>0$.
Since the matrix $I-N^{-1}H$ is invertible, we can express the matrix $Y$ in terms of $X$
using the first condition of (\ref{Yconstraints}),
\begin{equation}
Y=-X[I+2F(I-(M-N)^{-1} H)].
\label{Y}
\end{equation}
The second condition of (\ref{Yconstraints}) is then satisfied for any $X$ provided that
(\ref{noisecondition}) holds.

In order to further simplify the matrix $Y$, we need to establish the relationship between 
$x_j$ and $n_j$. Without loss of generality, we can
restrict ourselves to the cloning machines that satisfy
(\ref{noisecondition}), and minimize the cost $C(n_j)$ under the constraint
(\ref{noisecondition}). Using the standard method of Lagrange multipliers,
we obtain the extremal equations for the optimal $n_j$'s for a given set of $x_j$'s,
\begin{equation}
x_j\sqrt{n_j}-\lambda(M-N)\sqrt{n_j}+\lambda \sum_{k=1}^M \sqrt{n_k}=0.
\label{extremalequations}
\end{equation}
with $\lambda$ being the Lagrange multiplier.
With the help of these formulas, we can show that
\begin{equation}
XF(I-(M-N)^{-1}H)=\frac{1}{\lambda(M-N)} XFX,
\label{Ysimplification}
\end{equation}
so that 
\begin{equation}
Y=-X - \frac{2}{\lambda(M-N)} XFX .
\label{Y2}
\end{equation}
The matrix $XFX$ is symmetric, hence $Y=Y^T$ as required.

The last step of the proof is to show that the matrix $Z$ is positive semidefinite. 
We first apply a transformation that preserves the positive semidefiniteness,
$\tilde{Z}=V Z V^\dagger$, where $V=\mathrm{diag}(X^{-1/2},X^{-1/2})$,
\[
\tilde{Z}=\left(\begin{array}{cc}
I & -iI -i2\eta X^{1/2}F X^{1/2} \\[1mm]
iI +i2\eta  X^{1/2}F X^{1/2} & I
\end{array}
\right),
\]
where $\eta=1/(\lambda(M-N))$.
We multiply  Eq.~(\ref{Y2}) with $X^{-1}$ and take the trace, so we find that
\begin{equation}
\eta \mathrm{Tr}[X^{1/2}F X^{1/2}]=
-1,
\label{normalization}
\end{equation}
where we made use of Eq. (\ref{noisecondition}). Since $F$ is  proportional to rank 
one projector the normalization (\ref{normalization}) implies 
that $\eta X^{1/2}F X^{1/2}=-|\phi\rangle\langle \phi|\equiv -\Phi$,
where $|\phi\rangle$ is a normalized real vector, $\langle \phi|\phi\rangle=1$. 
We can convert $\tilde{Z}$ to block diagonal form with the unitary (\ref{Umatrix}),
$U \tilde{Z} U^\dagger=\mathrm{diag}(2\Phi,2I-2\Phi)$,
which is obviously positive semidefinite. This concludes the proof of optimality of the
cloning machine (\ref{cloningtransformation}).

\section{Conclusions}

In summary, we have proposed a multipartite asymmetric Gaussian cloning machine
for coherent states. The machine produces $M$ approximate copies of a 
coherent state from $N$ replicas of this state 
in such a way that each copy can have a different fidelity. A simple analytical
formula characterizing the set of optimal Gaussian asymmetric cloning machines has been
derived, and it was shown that the asymmetric cloning can be realized by amplifying
of a part of the input signal followed by mixing the amplified signal and the bypass signal 
together with auxiliary vacuum modes on an array of beam
splitters with carefully chosen transmittances. An alternate implementation is also
described, where the amplifier is replaced by a passive optical circuit
supplemented with feedforward. We hope that our study of multipartite 
asymmetric cloning will trigger further investigations of optimal 
quantum information distribution in continuous-variable quantum communication networks.

\acknowledgments

We acknowledge financial support from the EU under projects 
COVAQIAL and SECOQC. JF acknowledges support from 
from the Ministry of Education of the Czech Republic
(LC06007 and MSM6198959213) and from GACR (202/05/0498).  
NJC acknowledges support from the IUAP programme of the Belgian government 
 under grant V-18.


\begin{thebibliography}{99}


\bibitem{Wootters82}
W.K. Wootters and W.H. Zurek, Nature (London) \textbf{299}, 802 (1982);
D. Dieks, Phys. Lett. \textbf{92A}, 271 (1982).

\bibitem{Gisin02}
N. Gisin, G. Ribordy, W. Tittel, and H. Zbinden,
Rev. Mod. Phys. \textbf{74}, 145 (2002).

\bibitem{Buzek96}
V. Bu\v{z}ek and M. Hillery, Phys. Rev. A \textbf{54}, 1844 (1996).

\bibitem{rev1} V. Scarani, S. Iblisdir, N. Gisin, and A. Ac\'\i n,
Rev. Mod. Phys. \textbf{77}, 1225 (2005).

\bibitem{rev2} N. J. Cerf and J. Fiur\' a\v sek, in: {\it Progress in Optics}, \textbf{49},
edited by E. Wolf, (Elsevier, Amsterdam, 2006), pp. 455-545.









\bibitem{Lamas-Linares02}
A. Lamas-Linares, C. Simon, J.C. Howell, and D. Bouwmeester,
Science {\bf 296}, 712 (2002).


\bibitem{Fasel02} 
 S. Fasel, N. Gisin, G. Ribordy, V. Scarani and H. Zbinden, 
 Phys. Rev. Lett. \textbf{89}, 107901 (2002). 
 
\bibitem{deMartini04}
F. DeMartini, D. Pelliccia and F. Sciarrino, 
Phys. Rev. Lett. \textbf{92}, 067901 (2004).
 
\bibitem{Ricci04}
 M. Ricci, F. Sciarrino, C. Sias and F. DeMartini, 
 Phys. Rev. Lett \textbf{92}, 047901 (2004).
  
\bibitem{Irvine04}
 W.T.M. Irvine, A. Lamas-Linares, M.J.A. de Dood, and D. Bouwmeester, 
 Phys. Rev. Lett \textbf{92}, 047902 (2004).


\bibitem{Masullo04}
L. Masullo, M. Ricci, and F. De Martini,  quant-ph/0412040.



\bibitem{Cerf98}
N. J. Cerf, Acta Phys. Slov. \textbf{48}, 115 (1998);
J. Mod. Opt. \textbf{47}, 187 (2000).



\bibitem{Niu98}
C.S. Niu and R.B. Griffiths, Phys. Rev. A \textbf{58}, 4377 (1998).


\bibitem{Cerf00}
N.J. Cerf, Phys. Rev. Lett. \textbf{84}, 4497 (2000). 





\bibitem{Zhao04}
Z. Zhao, A.-N. Zhang, X.-Q. Zhou, Y.-A. Chen, C.-Y. Lu, A. Karlsson, and J.-W. Pan,  
Phys. Rev. Lett. \textbf{95}, 030502 (2005).


\bibitem{Filip04}
R. Filip, Phys. Rev. A \textbf{69}, 032309 (2004); 
Phys. Rev. A \textbf{69}, 052301 (2004).



\bibitem{Iblisdir04}
 S. Iblisdir, A. Ac\'{\i}n, N.J. Cerf, R. Filip, J. Fiur\'{a}\v{s}ek, and N. Gisin,
Phys. Rev. A \textbf{72}, 042328 (2005).


\bibitem{multiasymm1} J. Fiur\'{a}\v{s}ek, R. Filip, and N. J. Cerf,
Quant. Inform. Comp. \textbf{5}, 585 (2005). 

\bibitem{multiasymm2} S. Iblisdir, A. Ac\'{\i}n, and N. Gisin,
e-print arXiv quant-ph/0505152.



\bibitem{Braunstein04}
S. L. Braunstein and P. van Loock, 
Rev. Mod. Phys. \textbf{77}, 513 (2005). 


\bibitem{Lindblad00}
G. Lindblad, J. Phys. A: Math. Gen. \textbf{33}, 5059 (2000).


\bibitem{Cerf00CV}
N.J. Cerf, A. Ipe, and X. Rottenberg,
Phys. Rev. Lett. \textbf{85}, 1754 (2000); 
N.J. Cerf and S. Iblisdir, Phys. Rev. A \textbf{62}, 040301 (2000). 


\bibitem{Braunstein01}
S.L. Braustein, N.J. Cerf, S. Iblisdir, P. van Loock, and S. Massar,
Phys. Rev. Lett. {\bf 86}, 4938 (2001).

\bibitem{Fiurasek01}
J. Fiur\' a\v sek, Phys. Rev. Lett. {\bf 86}, 4942 (2001).



\bibitem{Fiurasek04}
J. Fiur\'{a}\v{s}ek, N.J. Cerf, and E.S. Polzik,  
Phys. Rev. Lett. \textbf{93}, 180501 (2004).


\bibitem{Cochrane04}
P. T. Cochrane, T. C. Ralph, and A. Dolinska,
Phys. Rev. A \textbf{69}, 042313 (2004).

\bibitem{Filip04CV}
R. Filip, J. Fiur\'{a}\v{s}ek, and P. Marek,
Phys. Rev. A \textbf{69}, 012314 (2004).

 
\bibitem{Andersen05}
U. L. Andersen, V. Josse, and G. Leuchs,
Phys. Rev. Lett. \textbf{94}, 240503 (2005). 

\bibitem{Koike06} 
S. Koike, H. Takahashi, H. Yonezawa, N. Takei, S. L. Braunstein, T. Aoki, and A. Furusawa,
 Phys. Rev. Lett. \textbf{96}, 060504 (2006).




\bibitem{Ferraro05}
A.  Ferraro and M.G. A. Paris,
Phys. Rev. A \textbf{72}, 032312 (2005).


 

\bibitem{Cerf04}
N.J. Cerf, O. Krueger, P. Navez, R.F. Werner, and M.M. Wolf,
Phys. Rev. Lett. \textbf{95}, 070501 (2005).

\bibitem{Grosshans02}
F. Grosshans and Ph. Grangier, Phys. Rev. Lett. {\bf 88}, 7902 (2002).

\bibitem{Grosshans03}
F. Grosshans F, G. Van Assche, R.M. Wenger, R. Brouri, 
N.J. Cerf, and P. Grangier, Nature (London) \textbf{421}, 238 (2003).

\bibitem{Grosshans04}
F. Grosshans and N. J. Cerf, Phys. Rev. Lett. \textbf{92}, 047905 (2004). 



\bibitem{Josse06}
V. Josse, M. Sabuncu, N. J. Cerf, G. Leuchs, and U. L. Andersen,
Phys. Rev. Lett. \textbf{96}, 163602 (2006).  


\bibitem{Bae06}
J. Bae and A. Ac\'{\i}n,
Phys. Rev. Lett. \textbf{97}, 030402 (2006). 

\bibitem{Andersen06}
U. L. Andersen, M. Sabuncu, R. Filip, and G. Leuchs,
Phys. Rev. Lett. \textbf{96}, 020409 (2006).


\bibitem{Banaszek01} 
K. Banaszek, Phys. Rev. Lett. {\bf 86}, 1366 (2001).

\bibitem{Mista05} 
L. Mi\v{s}ta, Jr., J. Fiur\'a\v{s}ek, and R. Filip, 
Phys. Rev. A {\bf 72}, 012311 (2005).

\bibitem{Mista06b}
L. Mi\v{s}ta, Jr. and J. Fiur\'a\v{s}ek,
Phys. Rev. A {\bf 74}, 022316 (2006).

\bibitem{Sacchi06} M. F. Sacchi, Phys. Rev. Lett. {\bf 96}, 220502 (2006).

\bibitem{Sciarrino06} 
F. Sciarrino, M. Ricci, F. De Martini, R. Filip, and L. Mi\v{s}ta, Jr., 
Phys. Rev. Lett. {\bf 96}, 020408 (2006).

\bibitem{Mista06} 
L. Mi\v{s}ta, Jr., Phys. Rev. A \textbf{73}, 032335 (2006).




\bibitem{Boyd96}
L. Vandenberghe and S. Boyd, SIAM Rev. \textbf{38}, 49 (1996).


\end{thebibliography}
\end{document}